\documentclass[11pt,a4paper]{article}

\usepackage{bm,amsfonts,amssymb,mathrsfs,amsmath,ascmac}
\usepackage{graphicx}
\usepackage{hyperref}
\usepackage{color}
\usepackage{tikz}
\usetikzlibrary{intersections, calc}

\newcommand{\rem}[1]{\textcolor{blue}{\bf[#1]}}

\newcommand{\bra}[1]{\langle #1 \rvert}
\newcommand{\ket}[1]{\lvert #1 \rangle}

\newcommand{\ol}{\overline}
\newcommand{\wt}{\widetilde}
\newcommand{\wh}{\widehat}
\newcommand{\cR}{\mathcal{R}}
\newcommand{\cT}{\mathcal{T}}
\newcommand{\cA}{\mathcal{A}}
\newcommand{\cC}{\mathcal{C}}

\newcommand{\cW}{\mathcal{W}}

\makeatletter

\@addtoreset{equation}{section}
\makeatother

\date{\today}

\begin{document}

\begin{titlepage}

\renewcommand{\thefootnote}{\fnsymbol{footnote}}

\begin{flushright}
 {\tt 
 HRI/ST/1504\\
 RIKEN-MP-112\\
 RIKEN-STAMP-7
 }
\\
\end{flushright}

\vskip9em

\begin{center}
 {\Large {\bf 
 Transport Process in Multi-Junctions of Quantum Systems
 }}

 \vskip5em

 \setcounter{footnote}{1}
 {\sc Taro Kimura$^{(a)}$}\footnote{E-mail address: 
 \href{mailto:taro.kimura@riken.jp}
 {\tt taro.kimura@riken.jp}} 
 and
 \setcounter{footnote}{2}
 {\sc Masaki Murata$^{(b)}$}\footnote{E-mail address: 
 \href{mailto:m.murata1982@gmail.com}
 {\tt m.murata1982@gmail.com}}

 \vskip2em

{\it 
 $^{(a)}$Theoretical Research Division, RIKEN Nishina Center, 
 Saitama 351-0198, Japan 
 \\ \vspace{.5em}
 $^{(b)}$ Harish-Chandra Research Institute, Chhatnag Road, Jhusi, Allahabad 211019, India
}

 \vskip3em

\end{center}

 \vskip2em

\begin{abstract}
We consider the junction of multiple one-dimensional systems and study how conserved currents transport at the junction.
To characterize the transport process, we introduce reflection/transmission
 coefficients by applying boundary conformal field theory.
We compute the reflection/transmission coefficients for some examples to derive the closed formulas. 
The formulas demonstrate spin-flip transport, where the spin
 polarization is flipped at the junction.
\end{abstract}

\end{titlepage}

\tableofcontents

\hrulefill

\setcounter{footnote}{0}


\section{Introduction}\label{sec:intro}

Recent development of nanotechnology allows us to build an electric
circuit in nanoscale, which involves quantum mechanical nature of
electrons.
To control such a nanoscale circuit, we need to investigate fundamental properties of quantum wire junctions. 
Among the theoretical studies of the quantum wire junction, Ref.~\cite{Nayak:1999zz} pointed out that the number
of connected wires interestingly affects the fixed point of the
renormalization group flow.
In this sense, the quantum wire junction gains
interests not only for engineering applications, but also for
fundamental theoretical aspects.
For this purpose, there have been a number of works based on conformal field theory (CFT).
This is because low-energy behavior of the wider class of
one-dimensional systems can be described as Tomonaga--Luttinger liquid
(TLL) through the bosonization scheme.
See, for example,~a textbook on this topic~\cite{Giamarchi:2003}.

Most works in this field are based on the TLL description of
one-dimensional systems, which is just $c = 1$ free boson CFT.
The $c=1$ CFT enables us to describe the $U(1)$ degree of freedom, which
corresponds to electric charge.
However, recent development of spintronics also demands us
to incorporate $SU(2)$ spin degree of freedom into such nanoscale
devices.
In this case, it is desirable to implement $SU(2)$ symmetry manifestly
in order to investigate the spin-dependent property at the junction.
Although the $c=1$ CFT can treat the spin 1/2 system,
corresponding to the $SU(2)_{k=1}$ Kac--Moody algebra, 
$c \neq 1$ is necessary for describing generic spins,
due to the identification of the Kac--Moody level $k$ with the
spin $s$ as $s = k/2$~\cite{Affleck:1985wb,Affleck:1987ch}.

In this paper we study transport process at the multi-junction of
one-dimensional systems that have Lie algebraic symmetries.
We work with an arbitrary multiplicity $M$ by generalizing
the previous works for $M=2$~\cite{Quella:2006de,Kimura:2014hva}.
See also~\cite{Bachas:2001vj}.
This junction plays a similar role to an impurity in the
one-dimensional system.
In fact, one can map
both the junction system and the impurity system into a
(1+1)-dimensional system with a boundary by using the folding
trick~\cite{Wong:1994np,Oshikawa:1996ww,Oshikawa:1996dj}.
In this picture the information about the junction is implemented into
the boundary state for the two dimensional system.
Using the boundary state, we shall define the transmission/reflection coefficient of conserved currents at the junction. 
In addition, to compute the coefficients explicitly, we shall construct a boundary state corresponding to the multi-junction of $SU(2)$-symmetric systems.
We shall compute both energy and spin-current reflection/transmission coefficients to investigate the spin-dependent property of the
junction.
In particular the spin transport shows an interesting behavior, namely, 
the spin-flipping process.

This paper is organized as follows.
In Sec.~\ref{sec:Transmission}, we formulate the transport process at the
multi-junction by generalizing the formulation for the multiplicity
$M=2$~\cite{Quella:2006de,Kimura:2014hva}.
We point out that the R-matrix, which characterizes the transport at the
junction, is not symmetric in general for $M>2$, while it is always symmetric when $M=2$.
In Sec.~\ref{sec:permutation}, we apply this formalism to the permutation
boundary condition, which is the simplest example to demonstrate the asymmetric R-matrix.
In Sec.~\ref{sec:cascade}, we study the transport with the coset-type boundary condition.
We shall propose the associated boundary state by generalizing that
shown in \cite{Quella:2002ct}.
The explicit computation of reflection/transmission coefficients shows that the current transport more strongly depends on the multiplicity $M$ than the energy transport.
We also discuss its application to the boundary entropy in
Appendix~\ref{sec:entropy}.
We conclude this paper in Sec.~\ref{sec:discussion} with some discussions.

\section{Reflection and Transmission Coefficients}\label{sec:Transmission}

\subsection{Multi-junction of Quantum Systems}
\label{sec:model}
In this paper, we shall consider the system with $M$ one-dimensional quantum systems connected at a point. 
Each quantum system is characterized by the following Hamiltonian densities in
the field theoretical limit:
\begin{align}
\mathcal{H}_i = \frac{1}{2\pi (k_i+ h_i^{\vee})} d_{AB}^i J^{i,A} J^{i,B}
 \label{Ham_Sugawara}
\end{align}
where $i=1,\ldots, M$ is the label of the quantum systems.  $J^{i,A}$ is the current taking values in the Lie algebra $\mathcal{A}_i$ and the index $A$ runs over $A=1,\ldots,\operatorname{dim}\mathcal{A}_i$. 
For the moment, we apply generic Lie algebras $\mathcal{A}_i$ rather
than $su(2)$.
$d^i_{AB}$ is the inverse of the Cartan--Killing form and $h_i^{\vee}$ is the dual Coxeter number of the algebra $\mathcal{A}_i$. 
The Fourier modes of the current $J^{i,A}$ satisfy the Kac--Moody algebra $\widehat{\mathcal{A}}_i$:
\begin{align}
 [j^{i,A}_m, j^{i,B}_n] 
 & = (f^i)^{A B}_{~~~C}~ j^{i,C}_{n+m} 
 + k_i \, m \, d^{i,A B} \,\delta_{m+n,0}
 \, ,
\end{align}
where $f^i$ is the structure constant of $\cA_i$ and $k_i$ is the level
of $\wh{\cA}_i$.

\begin{figure}[t]
 \begin{center}
 \begin{tikzpicture}
  

  \draw [thick,->] (-8,0) -- ++ (-3,0) node [below] {$x_1$};
  \draw [thick,->] (-8,0) -- ++ (30:3) node [right] {$x_2$};
  \draw [thick,->] (-8,0) -- ++ (-35:3) node [right] {$x_3$};


  \filldraw [very thick,fill=white] (-8,0) circle [radius=0.1];
  \filldraw (-8.8,0) circle [radius=0.1];
  \filldraw (-9.6,0) circle [radius=0.1];
  \filldraw (-10.4,0) circle [radius=0.1];
  \filldraw ($(-8,0)+(30:0.8)$) circle [radius=0.1];
  \filldraw ($(-8,0)+(30:1.6)$) circle [radius=0.1];
  \filldraw ($(-8,0)+(30:2.4)$) circle [radius=0.1];
  \filldraw ($(-8,0)+(-35:0.8)$) circle [radius=0.1];
  \filldraw ($(-8,0)+(-35:1.6)$) circle [radius=0.1];
  \filldraw ($(-8,0)+(-35:2.4)$) circle [radius=0.1];



  \draw [thick,->] (-8,2.5) node [above] {Junction at $x_i=0$} -- (-8,0.5) ;


  \draw [thick,->] (0,0) -- (-3,0) node [below] {$x_1$};
  \draw [thick,->] (0,0) -- ++ (30:2.5) node [right] {$x_2$};
  \draw [thick,->] (0,0) -- ++ (-35:2.5) node [right] {$x_3$};


  \draw [thick,->] (-3,2.8) -- (-3,3.3) node [above] {$t$};


  \draw [thick] (0,0) -- (-2.5,0) -- (-2.5,3.) -- (0,3.) -- cycle;
  \draw [thick] (0,0) -- ++ (30:2) -- ++ (0,3.) -- (0,3.) --cycle;
  \filldraw [thick,fill=white,opacity=.7,draw=black] 
  (0,0) -- ++ (-35:2.3) -- ++ (0,3.) -- (0,3.) --cycle;
  \draw [thick] (0,0) -- ++ (-35:2.3) -- ++ (0,3.) -- (0,3.) --cycle;


  \draw [ultra thick] (0,0) -- (0,3.);


  \draw [->,very thick] (-0.5,-1) node [below] {$x_i = 0$} -- (-0.1,-0.2);

 \end{tikzpicture}
\end{center}
 \caption{The junction of one-dimensional systems with the multiplicity
 $M=3$. (Left) Each system is defined through the axis of $x_i>0$, and
 interacting with each other at the sharing origin $x_i=0$. (Right) By
 adding the time direction, one obtains several two-dimensional planes glued along the defect line at
 $x_i = 0$.}
 \label{fig:Y-junction}
\end{figure}
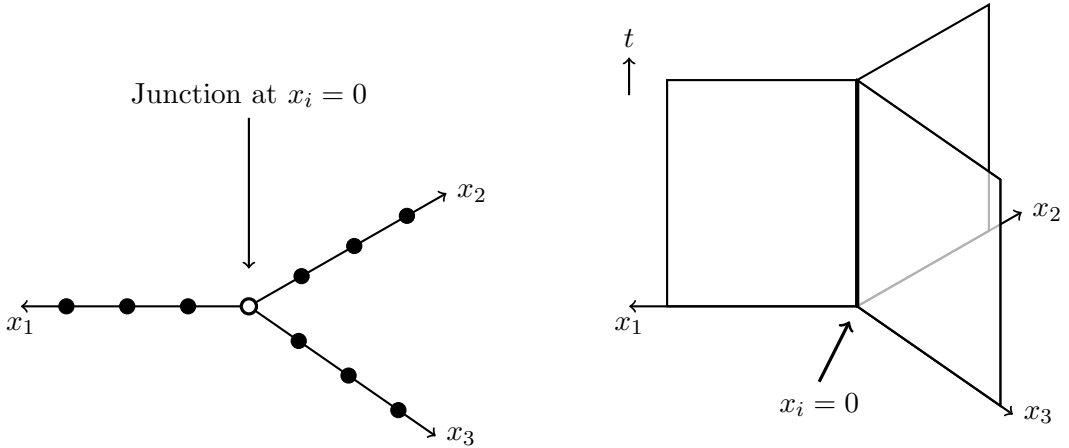

In addition to these Hamiltonians, we also introduce the
junction 
which connects the quantum systems through a local interaction. 
The interaction occurs if the ``spin'' $S^a$ at the junction takes value in a
subalgebra $\cC$ of $\cA_i$.
One possible local interaction is described by
\begin{align}
\mathcal{H}_i^\text{int} = \lambda_i \delta(x_i) d_{ab} J^{i,a} S^b \, ,
 \label{Ham_int}
\end{align}
where $x_i \geq 0$ is the coordinate of the quantum systems and the
junction is at $x_i=0$, as shown in Fig.~\ref{fig:Y-junction}.
The index $a$ runs as $a=1,\ldots,\operatorname{dim}\mathcal{C}$.

\begin{figure}[t]
\begin{center}
 \begin{tikzpicture}
 
 \filldraw [thick,fill=white,draw=black] 
 (0,0) -- (4,0) -- ++ (70:2) node [above] {CFT$_3$} -- ++ (-4,0) -- cycle;

 \filldraw [thick,fill=white,draw=black] 
 (0,0) -- (4,0) -- ++ (90:2.3) -- ++ (-2,0) node [above] {CFT$_2$} 
  -- ++ (-2,0) -- cycle;

 \filldraw [thick,fill=white,draw=black] 
 (0,0) -- (4,0) -- ++ (115:2) -- ++ (-4,0) node [above] {CFT$_1$} -- cycle;

 \draw [ultra thick] (0,0) -- (4,0);

  \draw [thick] (7.5,0) -- ++ (5,0) -- ++ (90:2.5) -- ++ (-5,0) -- cycle;
  \draw [ultra thick] (7.5,0) -- ++ (5,0);

  \draw [thick,->] (9.5,-0.5) -- ++ (1,0) node [right] {$t$};
  \draw [thick,->] (7.,1.8) -- ++ (0,0.5) node [above] {$x$};

  \node at (10,1.25) {CFT$_1 \times$CFT$_2 \times$CFT$_3$};

  \node at (5.8,-1) {$x=0$};
  \draw [very thick,->] (6.4,-0.8) -- (7.2,-0.2);
  \draw [very thick,->] (5.1,-0.8) -- (4.3,-0.2);

 \end{tikzpicture}
\end{center}
 \caption{A system with the conformal defect (left) is mapped into another one
 (right) with the boundary at $x=0$ through the folding trick. For example, a
 non-trivial bound sate at the junction can be studied using the BCFT
 approach.}
 \label{fig:folding_trick}
\end{figure}
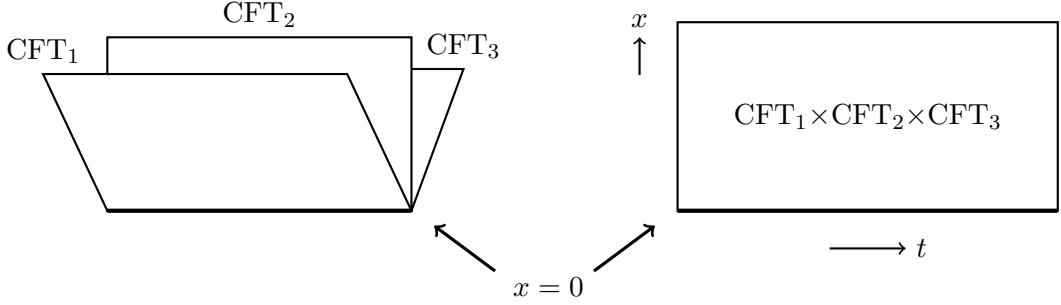

The critical point of this system is $M$-sheeted CFT glued along the
conformal defect corresponding to the world line of the junction. 
This configuration leads to ${\rm CFT}_1\times {\rm CFT}_2 \times
\cdots \times {\rm CFT}_M$, using the so-called folding
trick~\cite{Wong:1994np,Oshikawa:1996ww,Oshikawa:1996dj}.
See Fig.~\ref{fig:folding_trick}.
In the following subsections, we shall define the
reflection/transmission coefficient through the boundary CFT (BCFT). 
The BCFT is characterized by the boundary states $\ket{B}$ which describe the boundary
conditions.
For example, the energy conservation law along the boundary leads to the gluing condition for the Virasoro generators
\begin{align}
\left( L^{\rm tot}_n - \ol{L}^{\rm tot}_{-n} \right) \lvert B \rangle = 0
\label{eq;gluingL}
\end{align}
with $L^{\rm tot}_n=\sum_{i=1}^M L^i_n$. 
On the other hand, the current conservation law needs more consideration.
If there is a common subalgebra $\cC \subset \cA_i$ for all $i$, we have
\begin{align}
\left( j^{{\rm tot}, a}_n + \ol{j}^{{\rm tot}, a}_{-n} \right) \lvert B \rangle = 0
\end{align}
where $j^{{\rm tot}, a}_n=\sum_{i=1}^M j^{i,a}_n$ takes value in $\cC$.%
\footnote{
This gluing condition is the special case of the generic form
\begin{align}
\left( 
 j^{a}_n + \Omega\left(\ol{j}^{a}_{-n}\right) 
\right) \lvert B \rangle = 0 \, .
 \label{glue_auto}
\end{align}
with an automorphism $\Omega$ preserving the energy-momentum tensor.
See, for example, \cite{Recknagel:2013BCFT}.
}
Furthermore, in general, a subset of $\cA_i$'s has a larger subgroup
$\cC'\supset \cC$.
Supposing that $\cA_{i=1,\ldots, l}$ contains $\cC'$, the gluing condition can be written as
\begin{align}
\sum_{i=1}^{l} \left( j^{i,\alpha}_n + \ol{j}^{i,\alpha}_{-n} \right) \lvert B \rangle = 0
\end{align}
where $j^{i,\alpha}$ takes value in $\cC'/\cC$.
On the other hand, if $\cA_i$ has no bigger common subalgebra with $\cA_{j\neq i}$'s, we have
\begin{align}
\left(
j^{i,\cA_i/C}_n + \ol{j}^{i,\cA_i/C}_{-n}
\right)
\lvert B \rangle = 0 
\label{eq:glueJAC}
\end{align}
Now the properties of the junction are encoded in the boundary state
$\lvert B \rangle$.

\subsection{The R-matrices}

To define the reflection/transmission coefficients, we first introduce
the R-matrix for the energy by generalizing that for $M=2$,
\begin{align}
R^{ij}_T 
 = \frac{ \langle 0 \rvert L_2^i \bar{L}^j_{2} \lvert B \rangle}
        {\langle 0 \rvert B \rangle} \, .
\end{align}
Due to the gluing condition for the total current \eqref{eq;gluingL}, the R-matrix satisfies the following constraints
\begin{align}
\sum_{i=1}^M R^{ij}_T = \frac{c_j}2 \,, \quad
\sum_{j=1}^M R^{ij}_T = \frac{c_i}2 \, .
\label{eq:ConstraintsT}
\end{align}
These conditions give $2M-1$ constraints for the matrix elements. 
As a consequence of these constraints, the R-matrix has
$M^2-2M+1=(M-1)^2$ degrees of freedom in total.
Notice that this reproduces the result of
$M=2$, which yields only one degree of
freedom~\cite{Quella:2006de,Kimura:2014hva}.
Let us introduce another basis to express the $(M-1)^2$ degrees of
freedom in the R-matrix,
\begin{align}
\omega^{\alpha\beta}_T = \frac{ \langle 0 \rvert \cW_2^\alpha \ol{\cW}^\beta_{2} \lvert B \rangle }{\langle 0 \rvert B \rangle}
\end{align}
where $\alpha,\beta=1,\ldots ,M-1$ and
\begin{align}
\cW^\alpha_m = \sqrt{\frac{2}{c_{\alpha+1}C_\alpha C_{\alpha+1}}} \left( c_{\alpha+1}\sum_{\beta=1}^\alpha L^\beta_m - C_\alpha L^{\alpha+1}_m \right)\,,~~
C_\alpha = \sum_{\beta=1}^\alpha c_\beta \, .
\label{eq:cW_nT}
\end{align}
It turns out that this $\cW^\alpha_{-2}\ket{0}$ forms an orthonormal basis,
\begin{align}
\langle 0 \rvert \cW_{2}^\alpha \cW_{-2}^\beta \lvert 0 \rangle 
 = \delta_{\alpha\beta} \, .
 \label{orthonormal_W}
\end{align}
Then, to express the R-matrix in terms of $\omega^{\alpha\beta}_T$, we introduce
the inverse transformation of \eqref{eq:cW_nT},
\begin{align}
L^i_m = \sum_{\alpha=1}^{M-1} A^{i\alpha}_T \cW^\alpha_m + A^{iM}_T L^{\rm tot}_m
\end{align}
with
\begin{gather}
A^{i\alpha}_T = c_i \sqrt{ \frac{c_{\alpha+1}}{2C_\alpha C_{\alpha+1}} } \,,~~(i\leq \alpha<M)\,,~~
A^{i+1,i}_T = - \sqrt{ \frac{c_{i+1}C_i}{2C_{i+1}} }\,, \\
A^{iM}_T = \frac{c_i}{C_M}\,,~~
A^{i\alpha}_T = 0\,,~~(i>M+1)
\end{gather}
Using the coefficients $A$, we get
\begin{align}
R^{ij}_T = \sum_{\alpha,\beta=1}^{M-1} 
 A^{i\alpha}_T A^{j\beta}_T \omega^{\alpha\beta}_T 
 + \frac{c_{\rm tot}}2 A^{iM}_T A^{jM}_T \, ,
\end{align}
where we have used
\begin{align}
\langle 0 \rvert \cW^\alpha_2 \ol{L}_2^{\rm tot} \lvert B \rangle = \langle 0
 \rvert L_2^{\rm tot} \ol{\cW}^\alpha_2 \lvert B \rangle =  0 \, .
\end{align}
It is worth to emphasize that the R-matrix is not symmetric in general,
while it is always symmetric for $M=2$. 
This is one of the outcomes of the fact that the currents can transmit
through more than one channel.

In the same way, we define the R-matrix for the currents $j^i$ taking a value in the algebra~$\cA_i$,
\begin{align}
R_J^{ij,AB} = -\frac{ \langle 0 \rvert j_1^{i,A} \bar{j}_1^{j,B} \lvert B \rangle }{\langle 0 \rvert B \rangle}
 \, .
\end{align}
If we restrict to the common subalgebra $\cC$, the symmetry guarantees
that the R-matrix is in the product form as
$R_J^{ij,ab}=d^{ab}R_J^{ij}$.
Without loss of generality, we can assume that no pair of $\cA_i$ has bigger common subalgebra $\cC'\supset\cC$. 
If such a $\cC'$ exists, we can focus on the subsector of the R-matrix associated with $\cC'/\cC$ and do the same procedure as below.
With this setup the matrix elements including the index of $\cA_i/\cC$ are
\begin{align}
R^{ij,A'B'}=k_id^{A'B'}\delta^{ij}  \,,~~
R^{ij,A'a}=R^{ij,aA'}=0
\end{align}
for any $i,j$.
And here $A'$ is the index of $\cA_i/\cC$ or $\cA_j/\cC$.
On the other hand, the R-matrix for $\cC$ satisfies the constraints given by replacing a pair
$(R_T^{ij}, c_i)$ with $(R_J^{ij},2k_i)$ in
\eqref{eq:ConstraintsT},
\begin{align}
\sum_{i=1}^M R^{ij}_J = k_j \,, \quad
\sum_{j=1}^M R^{ij}_J = k_i \, . 
\label{eq:ConstraintsJ}
\end{align}
We can now utilize the same argument to obtain
\begin{align}
 R^{ij}_J = \sum_{\alpha,\beta=1}^{M-1} 
 A^{i\alpha}_J A^{j\beta}_J \omega^{\alpha\beta}_J 
 + k_{\rm tot} A^{iM}_J A^{jM}_J
\end{align}
using the orthonormal basis
\begin{gather}
\omega^{\alpha\beta}_J d^{ab} 
= -
\label{eq:omegaJ}
\frac{ \langle 0 \rvert \hat{K}_1^{\alpha,a} \ol{\hat{K}}^{\beta,b}_{1} \lvert B \rangle }
{\langle 0 \rvert B \rangle}
\\
\hat{K}^\alpha_m = \sqrt{\frac{1}{k_{\alpha+1}\kappa_\alpha \kappa_{\alpha+1}}} K^\alpha_m\,, ~~
K^\alpha_m = \left( k_{\alpha+1}\sum_{\beta=1}^\alpha j^\beta_m  - \kappa_\alpha j^{\alpha+1}_m \right) \,,~~
\kappa_\alpha = \sum_{\beta=1}^\alpha k_\beta \, ,
\label{eq:cK_nJ}
\end{gather}
with
\begin{gather}
A^{i\alpha}_J = k_i \sqrt{ \frac{k_{\alpha+1}}{\kappa_\alpha \kappa_{\alpha+1}} } \,,~~(i\leq \alpha<M)\,,~~
A^{i+1,i}_J = - \sqrt{ \frac{k_{i+1}\kappa_i}{\kappa_{i+1}} }\,, \\
A^{iM}_J = \frac{k_i}{\kappa_M}\,,~~
A^{iM}_J = 0\,,~~(i>n+1) \, .
\end{gather}

\subsection{Reflection and Transmission Coefficients}
\label{eq:def_trans}
Now we shall define the reflection and transmission coefficients using the
R-matrices defined above. 
As in Refs.~\cite{Quella:2006de,Kimura:2014hva}, it is natural to relate the diagonal and off-diagonal elements of the R-matrix to the reflection and transmission coefficients respectively. 
For the simplest case $M=2$ \cite{Quella:2006de,Kimura:2014hva}, the transmission rate is defined by the ``average'' of the off-diagonal elements since $R^{12}=R^{21}$, which is derived from the conservation law.
However, for $M>2$, the conservation law cannot give such a strong
constraint, and thus the average is not suitable to characterize the
transport process.
Therefore, we set $\cT^{ij}$ the transmission coefficient from system $i$ to
$j$ with $i\neq j$, and we shall treat  $\cT^{ij}$ and $\cT^{ji}$ as
independent variables.
Physically it is plausible to demand 
\begin{align}
1 = \cR^i_T + \sum_{j\neq i}\cT^{ij}_T \, , \quad
1 = \cR^i_J + \sum_{j\neq i}\cT^{ij}_J \, ,
\label{eq:RT1}
\end{align}
where $\cR^i$ is the reflection coefficient for the $i$-th system.
The constraints \eqref{eq:ConstraintsT} and \eqref{eq:ConstraintsJ} lead us
to define
\begin{align}
\cT^{ij}_T &= \frac{2}{c_i} R^{ij}_T\,,~~
\cT^{ij}_J = \frac{1}{k_i} R^{ij}_J\,,~~
\nonumber \\
\cR^{i}_T &= \frac{2}{c_i} R^{ii}_T\,,~~
\cR^{i}_J = \frac{1}{k_i} R^{ii}_J\,.~~
\end{align}
Here $R^{ij}_J$ is the R-matrix restricted to the
subalgebra $\cC$, assuming that no pair of $\cA_i$'s have a
bigger common subalgebra.
For $\cA_i/\cC$, it is plausible to set $\cR^{i}_J=1$ and $\cT^{ij}_J=0$
due to the gluing condition \eqref{eq:glueJAC}.

This definition does not reduce to the previous ones for $M=2$. 
However, since the R-matrix is symmetric for $M=2$, we have
\begin{align}
\mathcal{T}_T^{12} = \frac{c_2}{c_1} \mathcal{T}_T^{21} = \frac{c_1+c_2}{c_1} \mathcal{T}_T^{\rm avr}\, .
\end{align}
where $\mathcal{T}_T^{\rm avr}$ is the transmission coefficient defined in \cite{Quella:2006de,Kimura:2014hva}. 
The similar relation holds for the current.
As stated above, the new definition can be naturally extended to $M > 2$. 

In the following two Sections, we shall compute the
reflection/transmission coefficients for two examples.


\section{Example I: Permutation Boundary Condition}
\label{sec:permutation}
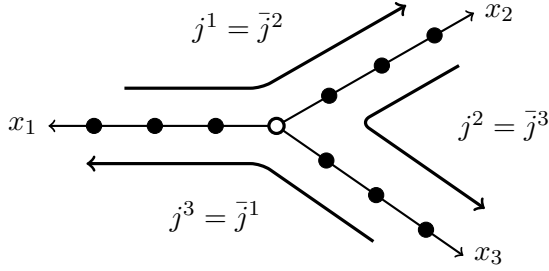
\begin{figure}[t]
\begin{center}
\begin{tikzpicture}


  \draw [thick,->] (-8,0) -- ++ (-3,0) node [left] {$x_1$};
  \draw [thick,->] (-8,0) -- ++ (30:3) node [right] {$x_2$};
  \draw [thick,->] (-8,0) -- ++ (-35:3) node [right] {$x_3$};


  \filldraw [very thick,fill=white] (-8,0) circle [radius=0.1];
  \filldraw (-8.8,0) circle [radius=0.1];
  \filldraw (-9.6,0) circle [radius=0.1];
  \filldraw (-10.4,0) circle [radius=0.1];
  \filldraw ($(-8,0)+(30:0.8)$) circle [radius=0.1];
  \filldraw ($(-8,0)+(30:1.6)$) circle [radius=0.1];
  \filldraw ($(-8,0)+(30:2.4)$) circle [radius=0.1];
  \filldraw ($(-8,0)+(-35:0.8)$) circle [radius=0.1];
  \filldraw ($(-8,0)+(-35:1.6)$) circle [radius=0.1];
  \filldraw ($(-8,0)+(-35:2.4)$) circle [radius=0.1];


 \draw [->,very thick] 
 (-10,0.5) [rounded corners] -- (-8.2,0.5) -- +(30:2.2);
 \draw [<-,very thick] 
 (-10.5,-0.5) [rounded corners] -- (-8.2,-0.5) -- + (-35:1.8);
 \draw [->,very thick]
 (-5.6,.8) [rounded corners] -- ++ (210:1.5) -- ++ (-35:2);


 \node at (-8.5,1.3) {$j^1 = \bar{j}^2$};
 \node at (-5.,0) {$j^2 = \bar{j}^3$};
 \node at (-8.8,-1.2) {$j^3 = \bar{j}^1$};

\end{tikzpicture} 
\end{center}
 \caption{The permutation boundary condition for $M=3$. All the currents
 for $\cC$ are transmitted to the next channel as described by the
 boundary condition (\ref{permutation_bc}).}
 \label{Permutation_fig}
\end{figure}

We first consider the case where the boundary condition is given by
\begin{align}
J^{i,a}(z) = \ol{J}^{i+1,a}(\bar{z})\,,~~
J^{i,\cA_i/\cC}(z) = \ol{J}^{i,\cA_i/\cC}(\bar{z})
 \label{permutation_bc}
\end{align}
with $J^{M+1}=J^1$ as shown in Fig.~\ref{Permutation_fig}.
Here $\cC$ is a subalgebra of all the $\cA_i$.
This gluing condition is consistent when all $k_i$ have the same value $k$. 
As stated in the previous section, $\cR^{i}_J=1$ for $\cA_i/\cC$. 
For $\cC$, we can straightforwardly compute
$R^{ij}_J$, and non-vanishing components are
\begin{align}
R_J^{12} = R_J^{23} = \cdots = R_J^{M1} = k \, .
\end{align}
It is interesting that the R-matrix is not symmetric for $M>2$. 
The transport coefficients for $\cC$ are
\begin{align}
\cR_J^{i} = 0\,,~~
\cT_J^{ij} = \delta_{i,j-1}\,.
\end{align}
This in fact satisfies the constraint \eqref{eq:RT1}. 

On the other hand, the R-matrix for the energy transport $R_T^{ij}$ is
more non-trivial,
\begin{align}
R_T^{ij} &= \frac{\langle 0 \rvert L_2^{i} \ol{L}_2^{j} \lvert B \rangle}{\langle 0 \rvert B \rangle}
= \frac{\langle 0 \rvert L_2^{i} \left( L_{-2}^{j-1,\cC} + L_{-2}^{j,\cA_j/\cC} \right) \lvert B \rangle}{\langle 0 \rvert B \rangle}
= \frac{c}2 \delta_{i,j-1} + \frac{c_i-c}2 \delta_{ij}
 \, .
\end{align}
where $c$ is the central charge corresponding to the algebra $\cC$ with level $k$. 
Thus the transport coefficients are
\begin{align}
\cR^i_T = 1 - \frac{c}{c_i} \,,~~
\cT^{ij}_T = \frac{c}{c_i} \delta_{i,j-1} \, .
\end{align}
Again it is easy to check \eqref{eq:RT1}.
Because $c_i>c$, we have $\cT^{ij}_T<1$. 
The physical interpretation of $\cT_J$ and $\cT_T$ is clear: 
currents for $\cC$ completely transmits with the contribution $c/c_i$
among the total energy, while the rest currents are
completely reflected giving $1-c/c_i$ contribution among the total energy.

\section{Example II: Coset-type Boundary Condition}
\label{sec:cascade}
Now we shall consider the $M$-junction of $SU(2)$ spin chains. 
To be more specific, we consider the cascade of breaking to the diagonal subgroup $SU(2)_{\kappa_M}$:
\begin{align}
& SU(2)_{k_1} \times SU(2)_{k_2} \times \cdots \times SU(2)_{k_M}
\nonumber\\
&\to 
\frac{SU(2)_{k_1} \times SU(2)_{k_2}}{SU(2)_{\kappa_2}} \times \frac{SU(2)_{\kappa_2}\times SU(2)_{k_3}}{SU(2)_{\kappa_3}} \times \cdots \times \frac{SU(2)_{\kappa_{M-1}}\times SU(2)_{k_M}}{SU(2)_{\kappa_M}} \times SU(2)_{\kappa_M}
 \, .
\end{align}
We claim that the corresponding boundary state is given by the following generalization of \cite{Quella:2002ct}:
\begin{align}
\lvert B(\rho,r) \rangle
&= \sum_{\mu,m}
\prod_{i=1}^M \frac{ S^{(k_i)}_{\rho_i \mu_i} }{ \sqrt{S^{(k_i)}_{0\mu_i}} } 
\prod_{i=1}^{M-1} \frac{ S^{(\kappa_{i+1})}_{r_i m_i} }{ S^{(\kappa_{i+1})}_{0m_i} } 
\lvert (\mu,m) \rangle \rangle
 \label{QS_type_state01}
\end{align}
where 
\begin{align}
\lvert (\mu,m) \rangle \rangle
= \lvert \mu_1,\mu_2,m_1 \rangle \rangle \otimes \lvert m_1,\mu_3,m_2 \rangle \rangle  \otimes 
\cdots \otimes \lvert m_{M-2},\mu_M,m_{M-1} \rangle \rangle  \otimes \lvert m_{M-1} \rangle \rangle 
\end{align}
is a product of $(M-1)$ Ishibashi states of each coset $SU(2)_{\kappa_i}\times SU(2)_{k_{i+1}}/SU(2)_{\kappa_{i+1}}$ and the Ishibashi state of $SU(2)_{\kappa_M}$. 
In Appendix \ref{sec:Cardy}, we shall show that this boundary state
satisfies the Cardy condition.
The other constraints on the boundary states, i.e. the sewing relations, are assumed. 

The parameters $(\rho,r)$ run over $2\rho_i=0,1,\cdots,k_i$,
$2r_i=0,1,\cdots,\kappa_{i+1}$, and $(\mu,m)$ runs over the same region
as $(\rho,r)$ satisfying the additional constraints:
\begin{align}
\mu_1+\mu_2+m_1 \in \mathbb{Z}\,,~~
m_1 + \mu_3+m_2 \in \mathbb{Z}\,,~~\cdots\,,
m_{M-2} + \mu_{M} + m_{M-1} \in \mathbb{Z} \, .
\label{eq:mum_region}
\end{align}
Not all the states labeled by $(\rho,r)$ are independent.
This is because the boundary state is invariant under 
\begin{align}
\rho_i \to J_{i-1,2} \rho_i\,,~~
r_i \to J_{i,3} J_{i+1,1}^{-1} r_i
\label{eq:BSidentify}
\end{align}
with $J_{0,2}=J_{1,1}, J_{M,1}=1$. 
Here, $(J_{i1},J_{i2},J_{i3})$ is an element of the identification group of $SU(2)_{\kappa_i}\times SU(2)_{k_{i+1}}/SU(2)_{\kappa_{i+1}}$. 
The two elements of that identification group can be expressed as
\begin{align}
(J_{i1}\mu_1,J_{i2}\mu_2,J_{i3}\mu_3) = (\mu_1,\mu_2,\mu_3) \,, \
 (\kappa_i/2-\mu_1,k_{i+1}/2-\mu_2,\kappa_{i+1}/2-\mu_3) \, ,
\end{align}
where $(\mu_1,\mu_2,\mu_3)$ labels a primary state of $SU(2)_{\kappa_i}\times
SU(2)_{k_{i+1}}/SU(2)_{\kappa_{i+1}}$. 

\subsection{Energy Transport}\label{sec:energy}

Recall that the R-matrix is defined by the overlap between $\langle 0
\rvert L_2^i \overline{L}_2^j$ and the boundary state $\ket{B}$, which
is now written in terms of coset states.
The descendant state $L_{-2}^i \lvert 0 \rangle$ shall be expanded by $SU(2)_{\kappa_M}$-singlet states with conformal weight $h=\bar{h}=2$,
\begin{align}
L_{-2}^i \lvert 0 \rangle 
 = \sum_{A=1}^{M(M+1)/2} \mathcal{L}_A^i \lvert \Sigma^A \rangle \, ,
\end{align}
where $\lvert \Sigma^A \rangle$ form a complete set of such singlet states. 
Here we shall use the explicit form $\{\ket{\Sigma^A}\}=\{\ket{V},\ket{W^\alpha},\ket{X^{\alpha,\beta}},\ket{Y^\alpha}\}$:
\begin{align}
\lvert V \rangle &= \lvert 0,0,0 \rangle_1 \otimes \cdots \otimes \lvert 0,0,0 \rangle_{M-1}  \otimes  L_{-2} \lvert 0 \rangle 
\nonumber \\
\lvert W^\alpha \rangle &= \lvert 0,0,0 \rangle_1 \otimes \cdots \otimes
 L_{-2} \lvert 0,0,0 \rangle_\alpha \otimes \cdots \otimes \lvert 0,0,0 \rangle_{M-1} \otimes \lvert 0 \rangle 
\nonumber \\
\ket{X^{\alpha,\beta}} &= K^\alpha_{-1} \cdot \wt{K}^\beta_{-1} \ket{0}
\nonumber \\
& \propto 
 \ket{0,0,0}_1 \otimes \cdots \otimes \ket{0,0,0}_{\alpha-1} \otimes \ket{0,0,1}_\alpha \otimes \ket{1,0,1}_{\alpha+1}
\nonumber \\
&\qquad\qquad
\otimes \cdots \otimes \ket{1,0,1}_{\beta-1} \otimes \ket{1,0,0}_\beta \otimes \ket{0,0,0}_{\beta+1} \otimes \cdots \ket{0,0,0}_{M-1} \otimes \ket{0} 
\nonumber \\
\ket{Y^\alpha} &= K^\alpha_{-1} \cdot j^{\rm tot}_{-1} \ket{0}
\nonumber \\
& \propto 
 \ket{0,0,0}_1 \otimes \cdots \otimes \ket{0,0,0}_{\alpha-1} \otimes \ket{0,0,1}_\alpha \otimes \ket{1,0,1}_{\alpha+1}
\otimes \cdots \otimes \ket{1,0,1}_{M-1} \otimes \ket{0}
\end{align}
with $\wt{K}^{\alpha,a}_n=k_{\alpha+1}\sum_{\beta=1}^\alpha j^{\beta,a}_n - (\kappa_{\alpha}+4) j^{\alpha+1,a}_{n}$.

Since the boundary state only has diagonal supports, we have
\begin{align}
R_T^{ij} 
&=  \sum_{A=1}^{M(M+1)/2}
 \mathcal{L}_A^i \mathcal{L}_A^j 
\frac{ \langle \Sigma^A \rvert \otimes \langle \Sigma^A | B \rangle} { \langle 0 \lvert B \rangle }
\nonumber \\
&= 
\sum_{A=1}^{M(M+1)/2}
 \frac{L_C(i,\Sigma^A) L_C(j,\Sigma^A)}{\langle \Sigma^A | \Sigma^A \rangle}
\frac{ \langle \sigma^A \rvert \otimes \langle \sigma^A \rvert B
 \rangle} { \langle 0 \lvert B \rangle } \, ,
\label{eq:RT-LC}
\end{align}
with the normalized state $\lvert \sigma^A \rangle =  \lvert \Sigma^A
\rangle/\sqrt{\langle \Sigma^A | \Sigma^A \rangle}$ and
\begin{align}
L_C(i,\Sigma^A) 
 = \langle \Sigma^A \rvert L_{-2}^i \lvert 0 \rangle 
 = \mathcal{L}_A^i \langle \Sigma^A | \Sigma^A \rangle
 \, .
\end{align}
This expression \eqref{eq:RT-LC} shows $R^{ij}_T = R^{ji}_T$; 
the boundary state \eqref{QS_type_state01} gives the symmetric R-matrix for energy. 
In the next subsection, we shall see that the spin-current R-matrix is also symmetric.

From now we shall compute $L_C$, $\langle \Sigma^A | \Sigma^A \rangle$
and $\bra{\sigma^A}\otimes\bra{\sigma^A}B\rangle/\bra{0}B \rangle$ one
by one.
The straightforward computation gives $L_C$ in the form of
\begin{align}
L_C(i,V) &= \frac{3k_i}{2(\kappa_M+2)}\,,
\nonumber \\
L_C(i,W^\alpha) &= 
\frac{3k_ik_{\alpha+1}}{2\left( \kappa_\alpha+2\right)\left(\kappa_{\alpha+1}+2\right)}U(\alpha-i) + \frac{3k_i \kappa_{i-1}}{2\left(k_i+2\right)\left( \kappa_i+2\right)}\delta _{\alpha+1,i}\,,
\nonumber \\
L_C\left(i,X^{\alpha,\beta}\right)
&= \frac{3}{2}k_{\alpha+1}k_{\beta+1}k_i U(\alpha-i)-\frac{3}{2}
 \kappa_\alpha k_{\beta+1}k_i\delta _{i,\alpha+1}\,,
\nonumber \\
L_C\left(i,Y^\alpha\right) &= \frac{3k_ik_{\alpha+1}}{2}U(\alpha-i) - \frac{3k_i \kappa_{i-1}}{2}\delta _{i,\alpha+1}
\label{eq:LCVWXY}
\end{align}
where $U(x)$ is a unit step function: $U(x\geq0)=1,U(x<0)=0$. 
We can also compute their norms as
\begin{align}
\langle V | V \rangle
&= \frac{c_{\kappa_M}}{2}\,,
\nonumber \\
\langle W^\alpha | W^\alpha \rangle
&= \frac{c_{\kappa_\alpha}+c_{\alpha+1}-c_{\kappa_{\alpha+1}}}{2}\,,
\nonumber \\
\langle X^{\alpha,\beta} | X^{\alpha,\beta} \rangle
&= \frac{3}{4} k_{\alpha+1}k_{\beta+1}\kappa_\alpha\kappa_{\alpha+1}\left(k_{\beta+1}\kappa_\beta+\left(\kappa_\beta+4\right){}^2+4k_{\beta+1}\right)\,,
\nonumber \\
\langle Y^\alpha | Y^\alpha \rangle
&= \frac{3}{4} \kappa_\alpha \kappa_{\alpha+1}k_{\alpha+1}\left(\kappa_M+4\right) \, .
\label{eq:normVWXY}
\end{align}
where $c_i$ and $c_{\kappa_i}$ are the center charges of $SU(2)_{k_i}$ and $SU(2)_{\kappa_i}$ respectively. 

For the boundary state \eqref{QS_type_state01}, we have
\begin{align}
\frac{\langle v\rvert \otimes \langle v|B\rangle }{\left<0|B\right>} &= 1\,,
\nonumber \\
\frac{\langle w^\alpha\rvert\otimes \left<w^\alpha|B\right> }{\left<0|B\right>} &= 1 \,,
\nonumber \\
\frac{\langle x^{\alpha,\beta}\rvert\otimes \left<x^{\alpha,\beta}|B\right> }{\left<0|B\right>} 
&= \left(\prod _{i=\alpha}^{\beta-1} \frac{S_{r_i1}^{\kappa_{i+1}}S_{00}^{\kappa_{i+1}}}{S_{01}^{\kappa_{i+1}}S_{r_i0}^{\kappa_{i+1}}}\right) \,,
\nonumber \\
\frac{\langle y^\alpha\rvert\otimes \left<y^\alpha|B\right>}{\left<0|B\right>} 
&= \left(\prod _{i=\alpha}^{M-1} \frac{S_{r_i1}^{\kappa_{i+1}}S_{00}^{\kappa_{i+1}}}{S_{01}^{\kappa_{i+1}}S_{r_i0}^{\kappa_{i+1}}}\right)\,.
\label{eq:vwxyB}
\end{align}
Inputting above data into \eqref{eq:RT-LC}, we obtain the closed
expression of the R-matrix. 
It is intriguing that the result is independent of $\rho_i$, which was
also for $M=2$ \cite{Quella:2006de,Kimura:2014hva}.
As in the case of the Kondo problem, the parameter $r_i$ could be
interpreted as the effective spin at the junction~\cite{Affleck:1995ge}.
This result implies the energy transport is basically characterized by
this residual effective spin of the junction.
As we will see below, this property is also observed for the current transport.

\subsection{Current Transport}\label{sec:current}

Let us now compute the reflection and transmission coefficients for Kac--Moody
current with the boundary state \eqref{QS_type_state01}. 
As shown below, the computation for the current is actually simpler than
that for the energy.
To compute the $\omega^{\alpha\beta}_J$ \eqref{eq:omegaJ}, it turns out
to be helpful to write $\wh{K}^{\alpha,+}_{-1} \ket{0}$ in terms of the
coset states.
$\wh{K}^{\alpha,+}_{-1}$ gives the descendants of
$SU(2)_{\kappa_\alpha}$ and $SU(2)_{k_{\alpha+1}}$, and makes the spin
$1$ state of $SU(2)_{\kappa_{\alpha+1}}$.
Thus we have
\begin{align}
\wh{K}^{\alpha,+}_{-1} \lvert 0 \rangle
= \lvert (\mu,m) \rangle ~~\text{with}~~
\mu_i=0\,,~~m_{i<\alpha}=0\,,~~m_{i\ge \alpha}=1 \, .
\end{align}
The relevant states to compute $R_J$ are the conformal vacuum $\ket{0}$
and the states with the conformal weight $h=\bar{h}=1$ and with spin $1$
under $SU(2)_{\kappa_M}$.
From \eqref{QS_type_state01}, these contributions are given by
\begin{align}
\lvert B(\rho,r) \rangle
= W_M\lvert 0 \rangle + \sum_{\alpha=1}^{M-1} W_\alpha \wh{K}^{\alpha,+}_{-1} \lvert 0 \rangle \widetilde{\wh{K}^{\alpha,-}_{-1} \lvert 0 \rangle} + \cdots
\end{align}
with
\begin{align}
W_j = \left( \prod_{i=1}^M \frac{ S^{(k_i)}_{\rho_i0} }{ \sqrt{S^{(k_i)}_{00}} }\right)
\left( \prod_{i=1}^{j-1} \frac{ S^{(\kappa_{i+1})}_{r_i 0} }{ S^{(\kappa_{i+1})}_{00} } \right)
\left( \prod_{i=j}^{M-1} \frac{ S^{(\kappa_{i+1})}_{r_i 1} }{ S^{(\kappa_{i+1})}_{01} } \right)
 \, .
\end{align}
Do not confuse this coefficient $W_j$ with the singlet state $\ket{W^\alpha}$
which appeared in Sec.~\ref{sec:energy}.
This leads to
\begin{align}
\omega^{\alpha\beta}_J = \delta_{\alpha,\beta} \frac{W_\alpha}{W_M}
= \delta_{\alpha,\beta}
\prod_{i=\alpha}^{M-1} \frac{ S^{(\kappa_{i+1})}_{00} S^{(\kappa_{i+1})}_{r_i 1} }{ S^{(\kappa_{i+1})}_{r_i 0} S^{(\kappa_{i+1})}_{01} }
\end{align}
and the R-matrix for the current
\begin{align}
R^{ij}_J 
 & =
 k_i k_j
 \left(
 -  \frac{1}{ K_{\alpha}} \frac{W_{\alpha-1}}{W_M}
 + \sum_{\beta=\alpha}^{M-1} 
 \frac{k_{\beta+1}}{K_\beta K_{\beta+1}} \frac{W_\beta}{W_M}
 + \frac{1}{K_M} 
 \right)
 \qquad \left( \alpha = \text{max}(i,j) \right)
 \, , \label{eq:RJSU2_01} \\
R^{ii}_J 
&= 
\frac{k_{i}K_{i-1}}{K_{i}} \frac{W_{i-1}}{W_M}
+ \sum_{\beta=i}^{M-1} \frac{k_i^2k_{\beta+1}}{K_\beta K_{\beta+1}} \frac{W_\beta}{W_M}
+ \frac{k_i^2}{K_M} \, .
\label{eq:RJSU2_02}
\end{align}
This expression shows that the spin-current R-matrix is symmetric and is independent of $\rho$'s as the energy R-matrix. 
Finally, the reflection and transmission coefficients for the currents are
\begin{align}
\cT^{ij}_J &= -  \frac{k_{j}}{ K_{\alpha}} \frac{W_{\alpha-1}}{W_M}
+ \sum_{\beta=\alpha}^{M-1} 
 \frac{k_j k_{\beta+1}}{K_\beta K_{\beta+1}} \frac{W_\beta}{W_M}
+ \frac{k_j}{K_M} 
 \quad \left( \alpha = \text{max}(i,j) \right) \,,
\\
\cR^{i}_J 
&= \frac{K_{i-1}}{K_{i}} \frac{W_{i-1}}{W_M}
+ \sum_{\beta=i}^{M-1} 
 \frac{k_ik_{\beta+1}}{K_\beta K_{\beta+1}} \frac{W_\beta}{W_M}
+ \frac{k_i}{K_M}
\,.
\end{align}
It is straightforward to show the conservation law:
\begin{align}
\sum_{j(\neq i)}^M \cT^{ij}_J + \cR^{i}_J = 1 \, .
\end{align}

\subsection{Results}

From now on, we shall show results for some multiplicity $M$ and parameters $r_i$.
First of all, let us check that our formula reproduces the previous
results for the simplest case $M=2$.
In this case, the Virasoro singlet states are
given by $\left\{\ket{\Sigma^A}\right\}=\{ \ket{V}, \ket{W^1},\ket{Y^1}
\}$, and thus we obtain
\begin{align}
R^{ij}_T &= 
 \frac{L_C(i,V) L_C(j,V)}{\langle V | V \rangle}
\frac{ \langle v \rvert \otimes \langle v \rvert B \rangle} { \langle 0 \lvert B \rangle }
+ \frac{L_C(i,W^1) L_C(j,W^1) }{\langle W^1 | W^1 \rangle}
\frac{ \langle w^1 \rvert \otimes \langle w^1 \rvert B \rangle} { \langle 0 \lvert B \rangle }
\nonumber \\
&+ \frac{L_C(i,Y^1) L_C(j,Y^1) }{\langle Y^1 | Y^1 \rangle}
\frac{ \langle y^1 \rvert \otimes \langle y^1 \rvert B \rangle} { \langle 0 \lvert B \rangle }
 \, .
\end{align}
By substituting \eqref{eq:LCVWXY}-\eqref{eq:vwxyB}, we reproduce the
result of Quella, Runnel and Watts~\cite{Quella:2006de} up to the conventions; $r_1$ in this paper is $\rho$ in \cite{Quella:2006de}.
In a similar way, for the current transport, the formulas
\eqref{eq:RJSU2_01} and \eqref{eq:RJSU2_02} reproduce our previous
result in \cite{Kimura:2014hva}.
Note that, as addressed in Sec.~\ref{eq:def_trans}, the definitions of
reflection/transmission coefficients are different, while $R_J$ and $R_T$ are
the same as the former definitions.

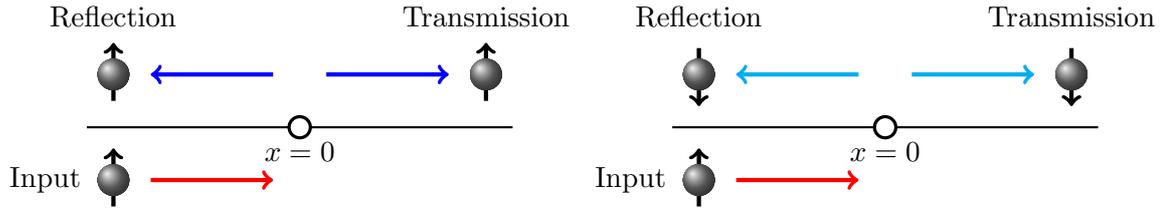
\begin{figure}[t]
\begin{center}
 \begin{tikzpicture}[scale=.7]
  

  \draw [thick] (-4,0) -- (4,0);
  \filldraw [fill=white,very thick] (0,0) circle (.2)
  node [below=.5mm] {$x=0$};


  \draw [->,ultra thick, red] (-2.8,-1) -- (-.5,-1);
  \draw [->,ultra thick, blue] (-.5,1) -- (-2.8,1);
  \draw [->,ultra thick, blue] (.5,1) -- (2.8,1);


  \begin{scope}[shift={(-3.5,-1)}]
   \draw [->,ultra thick] (0,-.5) -- (0,.6);
   \shade[ball color = gray] (0,0) circle (0.3) node [left=3mm] {Input};
  \end{scope}


  \begin{scope}[shift={(-3.5,1)}]
   \draw [->,ultra thick] (0,-.5) -- (0,.6);
   \shade[ball color = gray] (0,0) circle (0.3) node [above=5mm] {Reflection};
  \end{scope}


  \begin{scope}[shift={(3.5,1)}]
   \draw [->,ultra thick] (0,-.5) -- (0,.6);
   \shade[ball color = gray] (0,0) circle (0.3) node [above=5mm] {Transmission};
  \end{scope}

  \begin{scope}[shift={(11,0)}]


  \draw [thick] (-4,0) -- (4,0);
  \filldraw [fill=white,very thick] (0,0) circle (.2)
   node [below=.5mm] {$x=0$};


  \draw [->,ultra thick, red] (-2.8,-1) -- (-.5,-1);
  \draw [->,ultra thick, cyan] (-.5,1) -- (-2.8,1);
  \draw [->,ultra thick, cyan] (.5,1) -- (2.8,1);


  \begin{scope}[shift={(-3.5,-1)}]
   \draw [->,ultra thick] (0,-.5) -- (0,.6);
   \shade[ball color = gray] (0,0) circle (0.3) node [left=3mm] {Input};
  \end{scope}


  \begin{scope}[shift={(-3.5,1)}]
   \draw [->,ultra thick] (0,.5) -- (0,-.6);
   \shade[ball color = gray] (0,0) circle (0.3) node [above=5mm] {Reflection};
  \end{scope}


  \begin{scope}[shift={(3.5,1)}]
   \draw [->,ultra thick] (0,.5) -- (0,-.6);
   \shade[ball color = gray] (0,0) circle (0.3) node [above=5mm] {Transmission};
  \end{scope}

  \end{scope}

 \end{tikzpicture}
\end{center}
 \caption{Schematic illustration of current transport process at the
 defect: (left) non-flipping and (right) spin-flipping processes.
 }
 \label{spin-flip_fig}
\end{figure}

The numerical computation with the new definition involves
Table~\ref{tab:M2} for $M=2$. 
The matrix, which we call the reflection/transmission matrix, in the
table is defined by
\begin{align}
U^{ij}_{T/J} & =
\begin{cases}
 \mathcal{R}^i_{T/J} & (\text{for} \ i=j) \\
 \mathcal{T}^{ij}_{T/J} & (\text{for} \ i \neq j)
\label{eq:RefTransMatrix}
\end{cases}
 \, .
\end{align}
Here we obtain negative transport coefficients for the current in
general. 
For example, if we apply the gluing condition \eqref{glue_auto} with the automorphism $\Omega=-1$
\begin{align}
 \Omega\left( \overline{j}_{-n}^{a} \right)
 = - \overline{j}_{-n}^{a}
 \, ,
 \label{auto_flip}
\end{align}
we immediately obtain the negative current transport coefficient.
This automorphism \eqref{auto_flip} implies redefinition of generators
of the corresponding Lie algebra.
In particular, for $SU(2)$, the redefinition of the generator $\sigma_z
\to -\sigma_z$ means the $z$-spin flip.
Therefore the negative transport can be understood as the flipping of the spin
polarization at the junction, as shown in Fig.~\ref{spin-flip_fig}.


\begin{table}[t]
\begin{center}
\begin{tabular}{ccc}
\hline \hline
$(k_1,k_2,r_1)$ &  $U_T$ & $U_J$ \\ \hline
$(1,3,1)$
&
$\left(
\begin{array}{cc}
 0.15625 & 0.84375 \\
 0.46875 & 0.53125 \\
\end{array}
\right)$
&
$\begin{pmatrix}
-0.125 & 1.125 \\
0.375 & 0.625
\end{pmatrix}$
\\ & & \\
$(1,5,3/2)$
&
$\begin{pmatrix}
 0.292893 & 0.707107 \\
 0.329983 & 0.670017
\end{pmatrix}$
&
$\begin{pmatrix}
-0.178511 & 1.17851 \\
0.235702 & 0.764298
\end{pmatrix}$
 \\
\hline \hline
\end{tabular}
\end{center}
\caption{Spin-flipping reflection coefficients for $M=2$. $U_{T/J}$ is the reflection/transmission matrix defined in \eqref{eq:RefTransMatrix}.}
\label{tab:M2}
\end{table}

\begin{table}[t]
\begin{center}
\scalebox{0.9}[0.9]{
\begin{tabular}{ccc}
\hline\hline
$(r_1,r_2,r_3)$ &  $U_T$ & $U_J$ \\ \hline
$(0,0,1)$ 
&
$
\left(
\begin{array}{cccc}
 0.15625 & 0.28125 & 0.28125 & 0.28125 \\
 0.28125 & 0.15625 & 0.28125 & 0.28125 \\
 0.28125 & 0.28125 & 0.15625 & 0.28125 \\
 0.28125 & 0.28125 & 0.28125 & 0.15625 \\
\end{array}
\right)
$
&
$
\left(
\begin{array}{cccc}
 -0.125 & 0.375 & 0.375 & 0.375 \\
 0.375 & -0.125 & 0.375 & 0.375 \\
 0.375 & 0.375 & -0.125 & 0.375 \\
 0.375 & 0.375 & 0.375 & -0.125 \\
\end{array}
\right)
$
\\ & & \\
$(1/2,0,1)$
&
$
\left(
\begin{array}{cccc}
 0.28125 & 0.15625 & 0.28125 & 0.28125 \\
 0.15625 & 0.28125 & 0.28125 & 0.28125 \\
 0.28125 & 0.28125 & 0.15625 & 0.28125 \\
 0.28125 & 0.28125 & 0.28125 & 0.15625 \\
\end{array}
\right)
$
&
$
\left(
\begin{array}{cccc}
 0.375 & -0.125 & 0.375 & 0.375 \\
 -0.125 & 0.375 & 0.375 & 0.375 \\
 0.375 & 0.375 & -0.125 & 0.375 \\
 0.375 & 0.375 & 0.375 & -0.125 \\
\end{array}
\right)
$ \\
\hline \hline
\end{tabular}
}
\end{center}
\caption{Spin-flipping reflection/transmission coefficients for $M=4$ with
 $k_i=1$ for $i=1,\ldots,4$, corresponding to the quadruple junction of
 the $s=1/2$ Heisenberg spin chains.}
\label{tab:M4}
\end{table}

We have observed the negative reflection coefficients even for $M=2$
when $k_1 + k_2 \geq 4$.
Due to the conservation law, not both of the transmission and the reflection coefficients can be negative. 
Our computation for $1\leq k_1,k_2 \leq 10$ shows that the transmission
coefficients are always positive. 
In addition, keeping $k_1=1$ fixed, the large $k_2$ gives the large absolute value of the negative reflection coefficient.
This implies that the spin chain with higher spins can flip a spin more efficiently. 
This property would be helpful for actual applications to control the
spin current.

For $M>3$, we obtain many examples of the spin-flipping process. 
In particular, for $M=4$, the spin-flipping is observed for $k_i=1$ with
$i=1,\ldots,4$, which is the quadruple junction of the $s=1/2$
Heisenberg spin chains, as shown in Table~\ref{tab:M4}.

In contrast to the current transport, it is expected that the energy
transport is always (semi-)positive, because it is not possible to
provide a reasonable interpretation for the negative energy transport.
Up to now, we do not know how to prove this for generic parameters. 
Even for $M=2$, the complete proof is missing in spite of an attempt
given in \cite{Quella:2006de}. 
For this reason, instead of giving a generic proof, we have done the
numerical tests for some values of parameters $M,r_i$.  
For $M=2, k_i\leq10$, $M=3, k_i\leq5$ and $M=4,k_i\leq2$, we have
explicitly observed the semi-positivity of the elements of $U_T$. 



\section{Discussion}\label{sec:discussion}

In this paper we have discussed the transport process at the
multi-junction with respect to both of energy and current flows.
We have defined the transport coefficients with arbitrary multiplicity
$M$ by modifying and generalizing the previous one for $M=2$.
We have applied this formalism to some examples.
The permutation boundary condition gives a simple, but important example
such that the transport process becomes asymmetric between the channels,
which cannot occur in the situation with $M=2$.
We have also considered the coset-type boundary condition, in order to
study the spin-dependent transport at the junction.
Proposing the corresponding boundary state, we have
seen the multiplicity-dependence of the transport coefficients.
By increasing the multiplicity, we have obtained more examples with the
negative current transport, while the energy transport coefficients are always positive.
This behavior suggests the spin-flip at the junction, which is more specific
to high-multiplicity.
In particular, for $M=4$, we have observed the spin-flipping reflection and
transmission even with the junction of the $s=1/2$ Heisenberg spin chains.
Thus the quadruple junction of $s=1/2$ Heisenberg spin chains seems
accessible both in experiments and theoretical studies.

Although the BCFT approach suggests a non-trivial fixed point involving
spin-flipping phenomena, it is still unclear what microscopic
interaction at the junction induces this phenomena.
In addition, the physical meaning of $\rho, r$ is not yet obvious.
As mentioned in Sec.~\ref{sec:energy}, a plausible interpretation of
$r_i$ is the effective spin at the junction, and this interpretation is
actually applied to the Kondo problem.
To understand them, it is important to analyze the multi-junction using
another method, e.g. Bethe ansatz.
In addition, for engineering applications, it is also required for
manipulating the boundary state.
A possible direction is to study the relation between the boundary state
and the interaction at the junction.
There are various choices of interaction terms; \eqref{Ham_int} is not
an unique choice.
A different interaction term may lead to a different boundary state. 
To understand how these interactions determine the boundary states is important
not only for actual applications, but also for theoretical interests.

In addition to the boundary states used in this paper,
there are a number of solutions to the boundary condition.
For the corresponding boundary states, it is interesting to study the transport coefficients.
For example, Fredenhagen and Quella \cite{Fredenhagen:2005an} proposed a
new type of boundary states, which is a generalization of
the permutation boundary state, used in Sec.~\ref{sec:permutation}.
However, the form of the boundary states is not well known.
One way to find them is to solve string field theory with
$SU(2)_{k_1}\times SU(2)_{k_2}$ symmetry.
As a first step, the authors are solving string field theory with the
single $SU(2)_k$ with collaborators. 

\subsection*{Acknowledgements}

MM thanks Ashoke Sen for useful discussions.
The work of TK is supported in part by Grant-in-Aid for JSPS
Fellows~(\#25-4302).
The work of MM is supported by JSPS postdoctoral fellowship for
research abroad (\#27-14).

\appendix
\section{Boundary entropy}\label{sec:entropy}

In addition to the transport process discussed in the main part of this
paper, another interesting application of the boundary state is the
boundary entropy, which is also called the 
$g$-factor~\cite{Affleck:1991tk}.
The boundary entropy is obtained by the inner product of the boundary state and the
conformal vacuum, with a proper subtraction of the ``bulk''
contribution,
\begin{equation}
 S_{\rm bdry} 
  = \ln \left< 0 | B \right> - \ln \left< 0 | B_0 \right> \, .
\end{equation}
This bulk contribution $S_0 = \ln \left< 0 | B_0 \right>$ corresponds to the
situation in the absence of the interaction between the junction and the
bulk.
Therefore $\ket{B_0}$ is given by
\begin{equation}
 \left| B_0 \right> =
  \left| 0 \right>^{\otimes M} \, ,
\end{equation}
where $\left| 0 \right>$'s are Cardy's boundary states for $SU(2)_{k_m}$ for $m=1,\ldots,M$.

As pointed out in the previous work~\cite{Kimura:2014hva}, 
$\left| 0 \right>^{\otimes M} $ is obtained by setting all the parameters to be zero in the
boundary state (\ref{QS_type_state01}),
\begin{equation}
 \left| B_0 \right> = 
   \left| B (0,0) \right> \, .
\end{equation}
Since the overlap between the boundary state and the conformal vacuum is
given by
\begin{equation}
 \left< 0 | B(\rho,r) \right>
  =
  \prod_{i=1}^M \frac{S^{(k_i)}_{\rho_i 0}}{\sqrt{S^{(k_i)}_{00}}}
  \prod_{i=1}^{M-1} \frac{S^{(\kappa_{i+1})}_{r_i 0}}{\sqrt{S^{(\kappa_{i+1})}_{00}}}
  \, ,
\end{equation}
we obtain the boundary entropy as follows,
\begin{equation}
 W_{\rm bdry}
  \equiv \exp \left( S_{\rm bdry} \right)
  = 
  \prod_{i=1}^M \frac{S^{(k_i)}_{\rho_i 0}}{S^{(k_i)}_{00}}
  \prod_{i=1}^{M-1} \frac{S^{(\kappa_{i+1})}_{r_i 0}}{S^{(\kappa_{i+1})}_{00}}
  \, .
  \label{bd_entropy_QS}
\end{equation}
Here $W_{\rm bdry}$ implies the ground state degeneracy for the
boundary, which is referred to the $g$-factor.
This expression can be directly applied to the spin-chain
junction under the identification of the Kac--Moody level $k_i$ with the
spin representation $s_i$ as $s_i =
k_i/2$
~\cite{Affleck:1985wb,Affleck:1987ch}.
It is interesting to check the formula
(\ref{bd_entropy_QS}) by studying the discrete lattice models with the
Bethe ansatz method.

\section{Cardy Condition}
\label{sec:Cardy}
Boundary states should satisfy consistency relations: the Cardy condition and the sewing relations. 
In this Appendix, we shall show that \eqref{QS_type_state01} satisfies the Cardy condition. 
While we have considered $M$ products of $SU(2)$ in Sec.~\ref{sec:cascade}, we shall treat $M$ products of a generic group $G$ in this Appendix. 
Now $\rho_i$ and $r_i$ are the weight of $G_{k_i}$ and $G_{\kappa_i}$
respectively. 
In the same way, the region of $(\mu,m)$ is specified by All$_G$ (see
eq.~(12) of Ref.~\cite{Quella:2002ct})
\begin{align}
(\mu, m) \in {\rm All}_G ~\Leftrightarrow~
\mu_1 + \mu_2 - m_1 \in \mathcal{G} \,,
m_1 + \mu_3 - m_2 \in \mathcal{G} \,,
\cdots\,,
m_{M-2} + \mu_M - m_{M-1} \in \mathcal{G}
\end{align}
with $\mathcal{G}$ the root lattice of $G$. 
This reduces to \eqref{eq:mum_region} when $G=SU(2)$.
Notice that the projection operator, which appears in
\cite{Quella:2002ct}, is trivial in this case. 

Now let us compute the partition function on the cylinder,
\begin{align}
Z_{(\rho,r),(\tau,t)} 
&= \bra{\rho,r} \wt{q}^{L_0+\bar{L}_0-c/12} \ket{\tau,t}
\nonumber \\
&= 
\sum_{(\mu,m)\in {\rm All}_G}
\left(\prod _{i=1}^M \frac{\bar{S}_{\rho _i\mu _i}^{k_i}}{\sqrt{S_{0\mu _i}^{k_i}}}\right)
\left(\prod _{i=1}^{M-1} \frac{S_{r_im_i}^{\kappa_{i+1}}}{S_{0m_i}^{\kappa_{i+1}}}\right)
\left(\prod _{i=1}^M \frac{S_{\tau _i\mu _i}^{k_i}}{\sqrt{S_{0\mu _i}^{k_i}}}\right) \left(\prod _{i=1}^{M-1} \frac{\bar{S}_{t_im_i}^{\kappa_{i+1}}}{S_{0m_i}^{\kappa_{i+1}}}\right)
\nonumber \\
&\qquad \qquad
\wt{\chi}_{\mu_1,\mu_2,m_1} \wt{\chi}_{m_1,\mu_3,m_2} \cdots \wt{\chi}_{m_{M-2},\mu_M,m_{M-1}}
\wt{\chi}_{m_{M-1}}
\end{align}
where $\wt{\chi}=\chi(\wt{q})$. Using the modular S-matrix, we get
\begin{align}
Z_{(\rho,r),(\tau,t)} 
&= |\mathcal{G}_{\rm id}|^{M-1} \sum_{\substack{(\mu,m)\in {\rm All}_G \\ [\nu,n,p]\in {\rm Rep}_G}}
\left( 
\prod _{i=1}^M  \frac{ \bar{S}_{\rho _i\mu _i}^{k_i} S_{\tau _i\mu _i}^{k_i} } { S_{0\mu _i}^{k_i} }
\right)
\left( 
\prod_{i=1}^{M-1} 
\frac{ S_{r_im_i}^{\kappa_{i+1}} \bar{S}_{t_im_i}^{\kappa_{i+1}} }
{ S_{0m_i}^{\kappa_{i+1}} S_{0m_i}^{\kappa_{i+1}} }
\right)
\nonumber \\
&\qquad \qquad
\left( \prod_{i=1}^M S^{k_i}_{\mu_i\nu_i} \right)
\left( \prod_{i=1}^{M-1} S^{\kappa_{i+1}}_{m_in_i} S^{\kappa_{i+1}}_{m_ip_i} \right)
\chi_{\nu_1,\nu_2,n_1}  \chi_{p_1,\nu_3,n_2} \cdots \chi_{p_{M-2},\nu_M,n_{M-1}} \chi_{p_{M-1}}
\nonumber \\
&=
|\mathcal{G}_{\rm id}|^{M-1} \sum_{\substack{(\mu,m)\in {\rm All}_G \\ [\nu,n,p]\in {\rm Rep}_G}}
\left( 
\prod _{i=1}^M  \frac{ \bar{S}_{\rho _i\mu _i}^{k_i} S_{\tau _i\mu _i}^{k_i} S^{k_i}_{\mu_i\nu_i}  } { S_{0\mu _i}^{k_i} }
\right)
\left( 
\prod_{i=1}^{M-1} 
\frac{ S_{r_im_i}^{\kappa_{i+1}} \bar{S}_{t_im_i}^{\kappa_{i+1}} }
{ S_{0m_i}^{\kappa_{i+1}} S_{0m_i}^{\kappa_{i+1}} }
\bar{S}^{\kappa_{i+1}}_{m_in_i} S^{\kappa_{i+1}}_{m_ip_i} 
\right)
\wh{\chi}_{[\nu,n,p]}
\end{align}
where $\chi=\chi(q)$ and we have used the modular transformation:
\begin{align}
\wt{\chi}_{m,n,p} = |\mathcal{G}_{\rm id}| \sum_{(m',n',p')\in {\rm Rep}_G} S^{k_1}_{mm'} S^{k_2}_{nn'} \bar{S}^{\kappa_1}_{pp'} \chi_{m'n'p'}
\end{align}
and $\wh{\chi}_{[\nu,n,p]}=\chi_{\nu_1,\nu_2,n_1}  \chi_{p_1,\nu_3,n_2} \cdots \chi_{p_{M-2},\nu_M,n_{M-1}} \chi_{p_{M-1}}$. 
$\mathcal{G}_{\rm id}$ is the identification group of $G\times G/G$ and $|\mathcal{G}_{\rm id}|$ is its dimension. 
Rep$_G$ is obtained by taking the quotient with respect to $\mathcal{G}_{\rm id}$ in each coset. 
For example,
\begin{align}
(\nu_1, \nu_2, n_1) \sim (J_{11}\nu_1, J_{12}\nu_2, J_{13} n_1)
\end{align}
where $(J_{11},J_{12},J_{13})\in \mathcal{G}_{\rm id}$. 
Now we can use the identity
\begin{align}
1 
= \frac1{|\mathcal{G}_{\rm id}|^{M-1} } 
\sum_{J}
e^{ 2\pi i
\left( Q_{J_{11}}(\mu_1) + Q_{J_{12}}(\mu_2) - Q_{J_{13}}(m_1) \right)
}
\left( 
\prod_{i=2}^{M-1} 
e^{2\pi i ( Q_{J_{i1}}(m_{i-1}) + Q_{J_{i2}}(\mu_{i+1}) - Q_{J_{i3}}(m_i) )
}
\right)
\end{align}
This holds if and only if $(\mu,m)\in$ Rep$_G$ otherwise the right hand side vanishes. 
By substituting this we get
\begin{align}
Z_{(\mu,r), (\tau,t)} &= 
\sum_{J,d} \sum_{ \substack{ \mu_i, m_i \in {\rm Rep}(G)  \\ [\nu,n,p]\in {\rm Rep}_G} }
e^{ 2\pi i
\left( Q_{J_{11}}(\mu_1) + Q_{J_{12}}(\mu_2) - Q_{J_{13}}(m_1) \right)
}
\left( 
\prod_{i=2}^{M-1} 
e^{2\pi i ( Q_{J_{i1}}(m_{i-1}) + Q_{J_{i2}}(\mu_{i+1}) - Q_{J_{i3}}(m_i) )
}
\right)
\nonumber \\
&\qquad\qquad
\left( 
\prod _{i=1}^M  \frac{ \bar{S}_{\rho _i\mu _i}^{k_i} S_{\tau _i\mu _i}^{k_i} S^{k_i}_{\mu_i\nu_i}  } { S_{0\mu _i}^{k_i} }
\right)
\left( 
\prod_{i=1}^{M-1} 
N_{r_i t_i^\dagger}^{\quad d_i} 
\frac{S_{d_i m_i}^{\kappa_{i+1}}} {S_{0m_i}}
\bar{S}^{\kappa_{i+1}}_{m_in_i} S^{\kappa_{i+1}}_{m_ip_i} 
\right) \wh{\chi}_{[\nu,n,p]}
\nonumber \\
&= 
\sum_{J,d} \sum_{ \substack{ \mu_i, m_i \in {\rm Rep}(G)  \\ [\nu,n,p]\in {\rm Rep}_G} }
\left( 
\prod _{i=1}^M  \frac{ \bar{S}_{\rho _i\mu _i}^{k_i} S_{\tau _i\mu _i}^{k_i}
e^{2\pi  i Q_{J_{i-1, 2}}\left(\mu _i\right) } S^{k_i}_{\mu_i\nu_i}  } { S_{0\mu _i}^{k_i} }
\right)
\nonumber \\
&\qquad
\left( 
\prod_{i=1}^{M-1} 
N_{r_i t_i^\dagger}^{\quad d_i} 
\frac
{S_{d_i m_i}^{\kappa_{i+1}}e^{-2\pi  i Q_{J_{i 3}}(m_i) }\bar{S}^{\kappa_{i+1}}_{m_in_i} 
e^{2\pi  i Q_{J_{i+1, 1}}(m_i)} S^{\kappa_{i+1}}_{m_ip_i} }
 {S_{0m_i}}
\right) \wh{\chi}_{[\nu,n,p]}
\nonumber \\
&= 
\sum_{J,d} \sum_{ \substack{ \mu_i,m_i{\rm Rep}(G)  \\ [\nu,n,p]\in {\rm Rep}_G} }
\left( 
\prod _{i=1}^M  \frac{ \bar{S}_{\rho _i\mu _i}^{k_i} S_{\tau _i\mu _i}^{k_i}
S^{k_i}_{\mu_i J_{i-1, 2}\nu_i}  } { S_{0\mu _i}^{k_i} }
\right)
\left( 
\prod_{i=1}^{M-1} 
N_{r_i t_i^\dagger}^{\quad d_i} 
\frac
{S_{d_i m_i}^{\kappa_{i+1}}\bar{S}^{\kappa_{i+1}}_{m_i J_{i 3}n_i} 
S^{\kappa_{i+1}}_{m_i J_{i+1, 1}p_i} }
 {S_{0m_i}}
\right) \wh{\chi}_{[\nu,n,p]}
\nonumber \\
&= 
\sum_{J,d} \sum_{ \substack{[\nu,n,p]\in {\rm Rep}_G} }
\left( \prod _{i=1}^M N_{\tau_i J_{i-1,2}\nu_i}^{~~~~~~~~~\rho_i} \right)
\left( 
\prod_{i=1}^{M-1} 
N_{r_i, t_i^\dagger}^{\quad d_i} 
N_{d_i, J_{i+1,1}p_i}^{\qquad\quad J_{i3}n_i}
\right) 
\wh{\chi}_{[\nu,n,p]}
\end{align}
with $J_{0,2}=J_{1,1}, J_{M,1}={\rm id}$. 
$N_{\mu\nu}^{~~\rho}$ is the fusion coefficient. 
Here we have used
\begin{align}
S_{J\mu\nu} = e^{2\pi i Q_J(\nu)} S_{\mu\nu}\,,~~
S_{\mu\nu} = S_{\nu\mu} \, .
\end{align}
Obviously, the coefficients of characters are semi-positive integers. 
In addition, the number of identity operator $(\nu=n=p=0)$ is
\begin{align}
\sum_{J,d} &
\left( \prod _{i=1}^M N_{\tau_i J_{i-1,2}0}^{~~~~~~~~~\rho_i} \right)
\left( 
\prod_{i=1}^{M-1} 
N_{r_i, t_i^\dagger}^{\quad d_i} 
N_{d_i, J_{i+1,1}0}^{\qquad\quad J_{i3}0}
\right) 
\nonumber \\
&=
\sum_{J,d} 
\left( \prod _{i=1}^M N_{\tau_i 0}^{~~~J_{i-1,2}^{-1}\rho_i} \right)
\left( 
\prod_{i=1}^{M-1} 
N_{r_i, t_i^\dagger}^{\quad d_i} 
N_{d_i, 0}^{\quad J_{i+1,1}^{-1} J_{i3}0}
\right) 
\nonumber \\
&= 
\sum_{J,d} 
\left( \prod _{i=1}^M \delta_{\tau_i}^{J_{i-1,2}^{-1}\rho_i} \right)
\left( 
\prod_{i=1}^{M-1} 
N_{r_i, t_i^\dagger}^{\quad d_i} \delta_{d_i}^{J_{i+1,1}^{-1} J_{i3}0}
\right) 
\nonumber \\
&= 
\sum_{J} 
\prod _{i=1}^M \delta_{\tau_i}^{J_{i-1,2}^{-1}\rho_i}
\prod_{i=1}^{M-1}  N_{r_i, t_i^\dagger}^{\quad J_{i+1,1}^{-1} J_{i3}0}
 \, .
\end{align}
Here we have used the following properties of the fusion matrix. 
\begin{align}
N_{\mu\nu}^{~~~\rho} = N_{\mu J\nu}^{~~~J\rho}\,,~~
N_{\mu0}^{~~~\rho} = \delta_\mu^\rho\,,~~
N_{\mu\nu}^{~~~\rho} = N_{\nu\mu}^{~~~\rho}\,,~~
N_{\mu\nu}^{~~~\rho} = N_{\mu\rho^\dagger}^{~~~\nu^\dagger} 
 \, .
\end{align}
The second factor can be transformed as
\begin{align}
N_{r_it_i{}^+}{}^{J_{i+1, 1}{}^{-1}J_{i,3}0} 
&= N_{r_i\left(J_{i+1, 1}{}^{-1}J_{i,3}0\right){}^+}{}^{t_i}
= \sum _{\sigma _i} \frac{\bar{S}_{J_{i+1, 1}{}^{-1}J_{i,3}0, \sigma _i}S_{r_i\sigma _i}S_{t_i\sigma _i}}{S_{0\sigma _i}}
\nonumber \\
&= \sum _{\sigma _i} e^{2\pi  i \left(Q_{J_{i+1,1}}\left(\sigma _i\right)-Q_{J_{i,3}}\left(\sigma _i\right)\right)}\frac{\bar{S}_{0, \sigma _i}S_{r_i\sigma _i}S_{t_i\sigma _i}}{S_{0\sigma _i}} 
\nonumber \\
&=\sum _{\sigma _i} \frac{S_{J_{i+1, 1}J_{i,3}{}^{-1}0, \sigma _i}S_{r_i\sigma _i}S_{t_i\sigma _i}}{S_{0\sigma _i}} 
= \sum _{\sigma _i} \frac{\bar{S}_{\left(J_{i+1, 1}J_{i,3}{}^{-1}0\right){}^+, \sigma _i}S_{r_i\sigma _i}S_{t_i\sigma _i}}{S_{0\sigma _i}}
\nonumber \\
&= N_{r_iJ_{i+1, 1}J_{i,3}{}^{-1}0}{}^{t_i} = N_{r_i0}{}^{J_{i+1, 1}{}^{-1}J_{i,3}t_i}\text{  }
= \delta _{r_i}^{J_{i+1, 1}{}^{-1}J_{i,3}t_i}\text{  }
\nonumber \\
&= \delta _{t_i}^{J_{i+1, 1}J_{i,3}{}^{-1}r_i}
 \, .
\end{align}
Finally, we get
\begin{align}
Z_{(\rho,r),(\tau,t)}
= 
\sum_{J} 
\prod_{i=1}^M \delta_{\tau_i}^{J_{i-1,2}^{-1}\rho_i}
\delta_{t_i}^{J_{i+1, 1}J_{i,3}{}^{-1}r_i} + \cdots
 \, .
\end{align}
The dots include the contribution from the other states. 
In the meantime, the $\ket{(\rho,r)}$ is invariant under \eqref{eq:BSidentify}.
Thus we conclude that the unique identity operator appears if and only if boundary states are equivalent up to $\mathcal{G}_{\rm id}$. 
This completes the proof of the Cardy condition.

\bibliographystyle{ytphys}
\bibliography{multi}

\end{document}